\documentclass[a4paper,12pt]{article}

\usepackage[dvipdfmx]{graphicx}
\usepackage{array}
\usepackage{amsmath}
\usepackage{amssymb}
\usepackage{latexsym}
\usepackage[normalem]{ulem}
\usepackage{cancel}
\usepackage{color}
\usepackage{ulem}
\usepackage{setspace}
\usepackage{cancel}
\usepackage{xcolor} 
\usepackage[dvipdfmx]{graphicx}

\newcommand{\Blue}[1]{\textcolor{blue}{#1}}

\date{empty}
\begin{document}
\begin{titlepage}
\null
\begin{flushright}
YGHP-16-07\\
TIT/HEP-657\\
Dec., 2016
\end{flushright}
\vskip 0.7cm
\begin{center}
{\Large \textbf{Thermal Inflation with Flaton Chemical Potential}}
\vspace{1cm}
\normalsize
\renewcommand\thefootnote{\alph{footnote}}

{\large
Masato Arai\footnote{masato.arai(at)yamagata-u.ac.jp},
Yoshishige Kobayashi$^\dagger $\footnote{yosh(at)th.phys.titech.ac.jp},
Nobuchika Okada $\ddagger$ \footnote{okadan(at)ua.edu}, \\
and Shin Sasaki$^\#$\footnote{shin-s(at)kitasato-u.ac.jp},
}
\vskip 0.7cm
  {\it
  Faculty of Science, Yamagata University \\ 
  Yamagata 990-8560, Japan \\
  \vskip 0.2cm
  $^\dagger$Department of Physics, Tokyo Institute of Technology \\
  Tokyo 152-8551, Japan \\
  \vskip 0.2cm 
  $^\ddagger$Department of Physics and Astronomy \\
  University of Alabama, Tuscaloosa, AL35487, USA \\
  \vskip 0.2cm 
  $^\#$Department of Physics,  Kitasato University \\
  Sagamihara 252-0373, Japan
}
\vskip 0.2cm
\begin{abstract}
Thermal inflation driven by a scalar field called ``flaton''
is a possible scenario to solve the cosmological moduli problem. We
study a model of thermal inflation with a flaton chemical potential. 
In the presence of the chemical potential, a negative mass squared of the 
flaton, which is necessary to terminate the thermal inflation, is naturally induced.
We identify the allowed parameter region for the chemical potential ($\mu$)
and the flaton self-coupling constant
to solve the cosmological moduli problem and satisfy theoretical
consistencies. In general, the chemical potential is a free parameter
and it can be taken to be much larger than the typical scale of soft
supersymmetry breaking parameters of $\mathcal{O} (1)$ TeV.
For $\mu \gtrsim 10^8$ GeV, we find that the reheating temperature after
the thermal inflation can be high enough for the thermal leptogenesis
scenario to be operative.
This is in sharp contrast to the standard thermal inflation scenario,
in which the reheating temperature is quite low and a special mechanism
is necessary for generating sufficient amount of baryon asymmetry in
the Universe after thermal inflation.
\vskip 0.5cm
\end{abstract}
\end{center}

\end{titlepage}

\newpage
\setcounter{footnote}{0}
\renewcommand\thefootnote{\arabic{footnote}}
\tableofcontents 
\pagenumbering{arabic}
\section{Introduction}
The exponentially accelerated expansion of spacetime in the early
period of the Universe is well-established as the cosmic inflation
scenario \cite{Sato:1980yn,Guth:1980zm, Starobinsky:1980te,Linde:1981mu, Albrecht:1982wi}.
The primordial inflation solves the flatness and the horizon problems in the Standard Big-Bang cosmology. 
On the other hand, supersymmetry (SUSY) is believed to play an important role 
in the study of elementary particles especially in the early stage of the Universe. 
It is known that the inflation scenarios in the supersymmetric epoch exhibit various problems.
Among other things, the relatively high reheating temperature after the primordial
inflation causes the overproduction of gravitino.
Late time decay of gravitino after the Big-Bang Nucleosynthesis deconstruct successfully synthesized light elements.
This is known as the gravitino problem \cite{Pagels:1981ke, Weinberg:1982zq, Krauss:1983ik}.
One resolution to the gravitino problem is achieved by a low reheating temperature  $T_{\rm RH}\lesssim 10^{6-7}$ GeV 
\cite{Kawasaki:2004yh, Kawasaki:2004qu}.

There is also a serious cosmological problem in the early Universe, known as the cosmological moduli 
problem \cite{Coughlan:1983ci, Banks:1993en, deCarlos:1993wie}.
The four-dimensional spacetime may be realized in superstring theories, which typically 
predict massless scalar excitations, {\it i.e.}, moduli fields.
Since the moduli fields only have Planck suppressed interactions, the energy density of the Universe
is dominated by the moduli fields before they decay.
If the moduli decay cannot reheat the Universe high enough $T_{\rm RH}\gtrsim 1$ MeV,
the present Universe cannot be realized.
This is the cosmological moduli problem.
This problem is intractable in the primordial inflation scenario
since the moduli particles are produced abundantly even in the low
reheating temperature.

In order to solve the moduli problem, a short period of the
secondary inflation
with $\mathcal{O}(10)$ e-foldings 
after the primordial inflation has been proposed \cite{Lyth:1995hj, Lyth:1995ka}.
By this second inflation, the number density of the moduli particles is diluted away
and their energy density never dominate the Universe.
Since this secondary inflation of spacetime is triggered by the thermal effect, 
this is called the thermal inflation.
Realization and phenomenological viability
of the thermal inflation have been discussed in detail, for example, 
in \cite{SKY, Barreiro:1996dx, Asaka:1997rv, Asaka:1999xd, Hiramatsu:2014uta, HMY}.

The thermal inflation is driven by a scalar field with an almost flat potential. 
This field is called flaton. 
The typical flaton potential at zero temperature is given by \cite{Lyth:1995ka} 
\begin{eqnarray}
 V (\phi) = V_0 - m_{\phi 0}^2 |\phi|^2 + \sum_{n=1}^{\infty} \lambda_n
  {|\phi|^{2n+4} \over \bar{M}_{\text{pl}}^{2n}}
\,, \label{pot-int}
\end{eqnarray}
where $\phi$ is the (complex scalar) flaton field,
$V_0$ is the vacuum energy at the origin, $m_{\phi 0}$ is the mass of
the flaton and $\lambda_n$ are the coupling constants. 
The higher dimensional interactions are suppressed by the reduced Planck
mass $\bar{M}_{\text{pl}}=2.4\times 10^{18}$ GeV.
Here the flaton is assumed to interact with a scalar
field $X$ which serves as the thermal bath, through which
the flaton potential $V$ receives finite temperature corrections from the thermal bath.
At a high temperature $T$, the effective mass squared $m^2 (T)$ of the
flaton behaves like $m^2 (T) = T^2 - m_{\phi 0}^2 > 0$ 
and the thermal inflation begins at $\phi = 0$. As the temperature decreases, the effective mass squared
turns negative, which leads to the violation of the slow-roll condition.
Therefore the tachyonic mass of the flaton is necessary for the end of the thermal inflation.
It has been discussed that the tachyonic mass is obtained by the renormalization group flow 
in a supersymmetric model \cite{Murayama:1992dj}. 
However, this does not happen in more general situations.
After the thermal inflation, the flaton rolls down to the true vacuum and then starts to oscillate there.
The flaton decays to the Standard Model particles to reheat the Universe.
This decay creates entropy, and the moduli problem can be solved.
In order to solve the moduli problem, the yield of the moduli field after the thermal inflation
must be reduced to $10^{-12}-10^{-15}$ \cite{Ellis:1990nb} or smaller.
However, this mechanism causes another problem: 
the entropy production by the flaton decay also dilutes the primordial baryon 
asymmetry produced by some mechanism beforehand. 
\footnote{This problem has been pointed out in the early stages of the development of the flaton field \cite{Yamamoto:1985mb}, 
before the proposal of the thermal inflation scenario.}
We need a mechanism to produce sufficient amount of baryon asymmetry before or after the thermal inflation. 
In \cite{SKY, Asaka:1999xd, HMY}, it has been studied whether sufficient baryon number asymmetry 
is produced with the use of the Affleck-Dine mechanism \cite{AD, DRT} after the thermal inflation. 
However, it was found that the Affleck-Dine mechanism is not phenomenologically viable in this framework. 
It is normally difficult to resolve the problem since the reheating temperature after the flaton decay 
is typically not high enough,
because of very weak couplings of the flaton to the Standard Model particles.

In this paper we propose a thermal inflation scenario that can solve the problems of termination of the thermal
inflation and of generating sufficient amount of
baryon asymmetry after the flaton decay.
For this purpose, we introduce a chemical potential $\mu$ for the flaton.
We will show that 
in the thermal effective potential, the chemical potential $\mu$ plays a role of the tachyonic mass 
of the flaton at low temperature. 
Hence, the thermal inflation ends when the chemical potential starts dominating over the thermal mass.
Furthermore, $\mu$ is a free parameter
in any system, which basically has nothing to do with soft SUSY breaking parameters.
This is in contrast with the standard thermal inflation scenario 
where the tachyonic mass term in (\ref{pot-int}) 
is supposed to be generated through SUSY breaking and hence we expect
 $|m_{\phi_0}| \simeq{\cal O}(1)$ TeV
for the weak scale SUSY.
The mass scale of the flaton is important since it determines the
reheating temperature ($T_{\rm RH2}$) after the flaton decay 
and what mechanism for the baryon number generation can be implemented.
In the standard thermal inflation scenario, $T_{\rm RH2}$ is at most
${\cal O}(100)$ MeV as we will discuss below.
With such a low reheating temperature, a possible scenario for
the baryon number generation is the Affleck-Dine mechanism 
\cite{AD, DRT}. 
As mentioned above, although the Affleck-Dine mechanism has been studied in 
models of the thermal inflation,
it turns out that sufficient amount of baryon number cannot be created
\cite{SKY, Asaka:1999xd, HMY}. 
In our model, we can set $\mu \gg 1$ TeV so that the reheating temperature can be much higher and
the thermal leptogenesis \cite{FY} (for review, see \cite{BDP}) can be operative 
even after the flaton decay.

The organization of this paper is as follows.
In the next section, we present a 
brief review on the standard thermal inflation.
In section 3, we introduce a chemical potential
for the flaton field and calculate the thermal effective potential of the flaton.
We then evaluate the yields of the moduli after the flaton decay and
identify
the allowed regions of the chemical potential
$\mu$ and the flaton coupling constant $\lambda$.
Section 4 is devoted to conclusions and discussions.
We give a brief derivation of the thermal effective potential in Appendix A.
In Appendix B, we derive the interaction term between the flaton and the Standard Model
gauge fields.

%
%
\section{Review of Standard Thermal Inflation}
In this section, we review the thermal inflation proposed in
\cite{Lyth:1995hj, Lyth:1995ka} and how the moduli problem is solved.  
If the flaton field causes thermal inflation,
the energy density by the oscillating moduli 
is diluted and hence the moduli problem can be solved.
After the thermal inflation,
the Universe is thermalized with the reheating temperature $T_{\rm RH2}$ through the flaton decay. 
If the reheating temperature is high enough to allow the Big-Bang nucleosynthesis ($T_{\rm RH2}\gtrsim 1$ MeV),
the history of the Universe becomes the standard scenario.

We focus on a part of a model which causes the thermal inflation while a part for the primordial inflation is not specified. 
We assume that the flaton field acquires its mass via SUSY breaking, and hence the mass is naturally of the order of the soft SUSY breaking mass scale $\sim 1$ TeV.
Notice that as we will see below,
the negative mass squared for the flaton field is necessary to 
terminate the thermal inflation.
For an origin of the negative mass squared, we may consider the renormalization group effect, which 
drives the running flaton mass squared negative at a certain low scale. 
For a concrete model, see \cite{Murayama:1992dj}.

The flaton field $\phi$ is considered to couple with some
light fields, typically the Standard Model particles, 
which are in thermal equilibrium and yield
thermal corrections to the effective potential of the flaton.
The high-temperature approximation is valid,
when the mass scale of the fields are sufficiently small compared to the temperature
during the thermal inflation.
For simplicity, we consider a model, 
with two real scalars $\phi$ and $X$
for the thermal inflation,
\begin{eqnarray}
 {\cal L}={1 \over 2}\partial_\mu \phi \partial^\mu \phi + {1 \over 2}\partial_\mu X \partial^\mu X -V(\phi,X)\,, \label{eq:lag}
\end{eqnarray}
where $\mu=0,1,2,3$ is the spacetime index, and we use the mostly minus
convention of the metric $\eta_{\mu\nu}={\rm diag}(1,-1,-1,-1)$. 
The scalar potential $V$ is given to be the following form
\begin{eqnarray}
 V_{\rm tree}=V_{0}-{m_{\phi 0}^2 \over 2} \phi^2+{\lambda \over 6\bar{M}_{\rm pl}^2}\phi^6+{m_{X0}^2 \over 2} X^2+{g\over 4}\phi^2 X^2\,, \label{eq:potential}
\end{eqnarray}
where $V_0$ is the energy scale at the origin, 
and $m_{\phi 0}$ and $m_{X0}$ are the masses of the fields $\phi$ and $X$,
respectively.
Here $\lambda$ and $g$ are coupling constants.
We have introduced the higher dimensional interaction term $\phi^6$,
and there is no flaton quartic term. \footnote{Such a form of the potential is found in low energy effective theory of superstring theories \cite{Dine:1985vv}. } 
This setting realizes an almost flat potential and leads to a large vacuum expectation value (VEV) of the flaton field.
\footnote{When the flat potential includes the flaton quartic coupling, it is necessary to set the coupling constant to be much smaller than $\lambda$ in (\ref{eq:potential}), in order to realize the large VEV.}
Such a large VEV is crucial to solve the moduli problem \cite{Lyth:1995ka} (see, (\ref{eq:yield-L}) with (\ref{eq:ph})).
The stationary condition for $X$ trivially gives $X=0$ while the one for $\phi$, 
\begin{eqnarray}
 {\partial V \over \partial \phi}{\Big |}_{X=0}=0
\end{eqnarray}
yields 
\begin{eqnarray}
\phi = 0, \qquad 
 \phi=\lambda^{-1/4}\sqrt{m_{\phi 0}\bar{M}_{\rm pl}}\equiv M\,. \label{eq:vacuum1}
\end{eqnarray}
The energy scale at the origin is given by
\begin{align}
 V_{0}={1 \over 3\sqrt{\lambda}}m_{\phi 0}^3\bar{M}_{\rm pl}={1 \over
 3}m_{\phi 0}^2M^2, \label{eq:ph}
\end{align}
which guarantees the vanishing cosmological constant at the potential
minimum $\phi_c = M$. We represent $\phi_c$ as the VEV of $\phi$.

The scalar potential (\ref{eq:potential}) receives thermal effects
through the reheating after the primordial inflation. The thermal
effects are introduced by imposing the periodic boundary condition for
the fields $\Phi_i=(\phi, X)$ as
$\Phi_i(\tau,\vec{x})=\Phi_i(\tau+\beta,\vec{x})$ in the partition
function, where $\tau=ix_0$ is the imaginary time, $\beta=1/T$ 
is the inverse temperature
and $\vec{x}=(x_1, x_2, x_3)$. The partition function is given as
\begin{eqnarray}
Z&=&{\rm Tr}e^{-\beta H} \nonumber \\
&=& \int_{\Phi_i(\tau)=\Phi_i(\tau+\beta)}
\prod_i
{\cal D}\Phi_i{\cal D}\Phi_i^\dagger e^{-\int_0^\beta d\tau \int d^3x \sum_i({1 \over 2}\partial_0 \Phi_i \partial_0 \Phi_i+{1 \over 2}\vec{\nabla} \Phi_i \vec{\nabla}\Phi_i+V(\phi, X))}\,,
\end{eqnarray}
where $H$ is the Hamiltonian and $\vec{\nabla}$ is 
the derivative with respect to $\vec{x}$. The scalar field $X$ 
plays the role of the thermal bath and the flaton receives the thermal
effects through $X$-loop corrections.
Calculating the thermal 1-loop correction of $X$,
we obtain the effective potential for the flaton as \sout{\cite{Actor}}
\Blue{
\footnote{For derivation, see Appendix A. }
}
\begin{eqnarray}
 V_{\rm eff}(\phi_c)&=&V_0-{1 \over 2}m_{\phi 0}^2 \phi_c^2+{\lambda \over 6\bar{M}_{\rm pl}^2} \phi_c^6+\int {d^3k \over (2\pi)^3}{\omega_k \over 2}+{1 \over \beta}\int {d^3 k \over (2\pi)^3} \log\left(1-e^{-\beta \omega_k}\right)\,, \label{eff-pot-T}
\end{eqnarray}
where we have defined
\begin{eqnarray}
 \omega_k^2&=&\vec{k}^2+ m_{X}^2(\phi) \,,  \label{eq:omega} \\
 m_X^2(\phi)&=&{\partial^2 V \over \partial X^2}{\Big |}_{\phi=\phi_c}=m_{X0}^2+{1\over 2}g\phi_c^2\,. \label{eq:mass-matrix} 
\end{eqnarray}
The fourth term in the right hand side of \eqref{eff-pot-T} is the
Coleman-Weinberg potential and the fifth term is the thermal effective potential.
We consider the situation where 
the temperature is high enough 
and the dominant contribution comes from the thermal effective
potential.
In the subsequent discussions, we therefore neglect the Coleman-Weinberg
potential term.
Performing the high temperature expansion, we have 
\begin{eqnarray}
 V_{\rm eff}(\phi_c)=V_0-{\pi^2 T^4 \over 90}+{T^2 \over 24}m_{X 0}^2+{1 \over 2}m_{\phi}^2(T)\phi_c^2+{\lambda \over 6\bar{M}_{\rm pl}^2}\phi_c^6+\cdots\,,
\end{eqnarray}
where $m_\phi(T)$ is the flaton mass
with the thermal correction:
\begin{eqnarray}
 m_\phi(T)^2=- m_{\phi 0}^2+{g \over 24}T^2\,.
\end{eqnarray}
For $m_\phi(T)^2>0$, the vacuum is located at $\phi_c=0$, and 
the potential energy of the flaton dominates over the energy of the Universe.
This leads to the second inflation by the flaton, namely, the thermal inflation.
The thermal inflation ends when the effective mass of the flaton turns
to be negative, in other words, when the temperature drops below the critical value
$T_C$ given by
\begin{eqnarray}
 T_C=2m_{\phi 0} \sqrt{6  \over g}\,. \label{eq:tc}
\end{eqnarray}
Soon after the temperature becomes less than $T_C$,
the flaton starts rolling down to the vacuum at $\phi_c=M$ and 
then oscillates around there.
The decay of the flaton reheats the Universe, and we roughly 
estimate the reheating temperature as
\begin{eqnarray}
 T_{\rm RH2} \simeq \left(90 \over \pi^2 g_*\right)^{1/4} \sqrt{\Gamma \bar{M}_{\rm pl}}\,, \label{eq:RH2}
\end{eqnarray}
where $g_*(\simeq 200)$ counts the effective degrees of freedom of the
radiation, and $\Gamma$ is the flaton decay width. Here we simply assume
that the flaton decays to the Higgs boson ($h$) through the effective interaction \cite{Asaka:1997rv} 
\begin{eqnarray}
 {\cal L}_{\rm int} \sim {m_{\phi}^2 \over M}\phi h h\,, \label{eq:int}
 \end{eqnarray}
where $m_{\phi}$ is the flaton mass 
in the vacuum at $T=0$ and given by
\begin{eqnarray} 
m_\phi^2=
{\partial^2 V \over \partial \phi^2}{\Big |}_{T=0, \phi_c=M}=4m_{\phi 0}^2\,.
\end{eqnarray}
The decay width of the process $\phi \rightarrow hh$ is obtained as
\begin{eqnarray}
 \Gamma \simeq {1 \over 16 \pi} {m_{\phi}^3 \over M^2}\,,\label{eq:gamma1}
\end{eqnarray}
where we have neglected the Higgs boson mass.
Substituting (\ref{eq:gamma1}) into (\ref{eq:RH2}), we 
find the reheating temperature as
\begin{eqnarray}
 T_{\rm RH2} \simeq  \left(90 \over \pi^2 g_*\right)^{1/4}{m_{\phi} \over 4M}\sqrt{m_{\phi} \bar{M}_{\rm pl}\over \pi}={1 \over 4\pi}\left(360\lambda \over g_*\right)^{1 \over 4}m_\phi\,. \label{eq:R-RH2}
\end{eqnarray}

\begin{figure}[tb]
\centering
\includegraphics[scale=0.9]{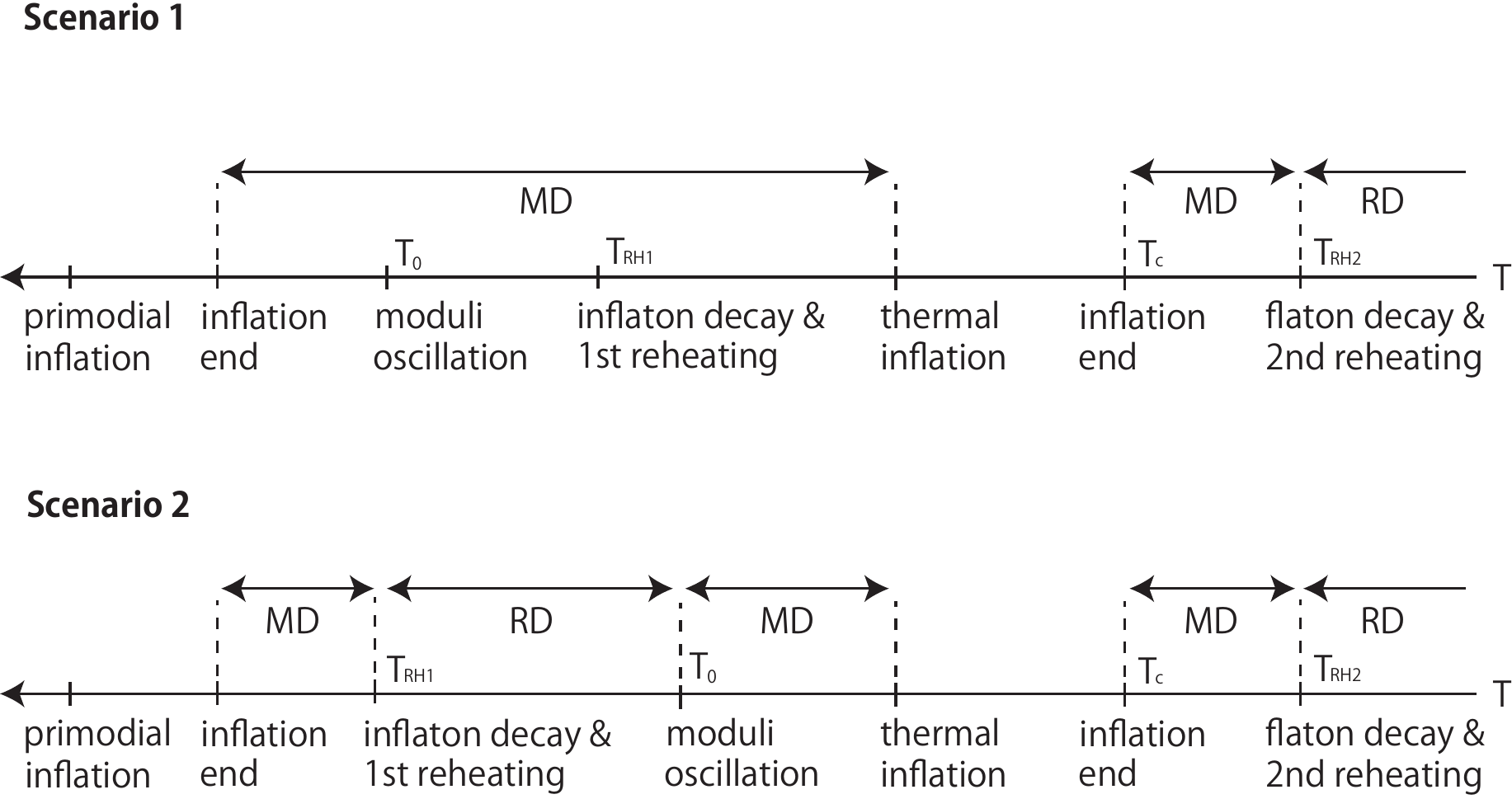}
\caption{
Two possible scenarios for the moduli oscillation in the thermal history
 of the Universe.
Here MD and RD mean the matter dominant and the radiation dominant, respectively.
}
\label{fig:TH}
\end{figure}

The main role of the thermal inflation is to dilute the yield of the
moduli field, by which the moduli problem is solved. The dilution is
caused by the entropy production by the flaton decay after the thermal
inflation. 
Before we discuss the entropy production, we note that
there are two relevant scenarios for the moduli oscillation after the
primordial inflation (see Fig. \ref{fig:TH}).
The first is the one discussed in \cite{Lyth:1995ka}.
In this scenario, the moduli fields are displaced from the potential
minima during the primordial inflation.
When the Hubble parameter reduces to $H \sim m_{\Phi}$, 
the moduli fields start to oscillate around
their potential minima. Here $m_\Phi$ is the mass of the moduli fields.
The Universe enters the matter dominated era with the oscillating
inflaton and moduli fields whose energy densities are comparable.
After the moduli oscillation, the first reheating takes place by the
decay of the inflaton and we denote the reheating temperature by $T_{\text{RH}1}$.

The second possibility is that the moduli oscillation takes place after
the first reheating. When the Universe cools down to $H\sim m_\Phi$, the
moduli fields start to oscillate.
As we will see later, in both scenarios, the oscillating moduli, that dominate the 
energy density of the Universe, can be diluted away by the thermal inflation.

In the following, we make a qualitative analysis on the entropy
production in these scenarios.

%
%

\paragraph{Scenario 1}
The increase of the entropy density after the flaton decay is calculated as
\begin{eqnarray}
\Delta={s(T_{\rm RH2}) \over s(T_C)}\,, \label{eq:increase}
\end{eqnarray}
where $s(T)$ is the entropy density at temperature $T$. 
In the radiation dominated era, this is given by 
\begin{eqnarray}
 s(T)={2\pi^2 \over 45}g_*T^3={4 \rho(T) \over 3 T}\,, \label{eq:entropy1}
\end{eqnarray}
where we have used the energy density for relativistic particles
\begin{eqnarray}
 \rho(T)={\pi^2 \over 30}g_*T^4\,. \label{eq:energy-density}
\end{eqnarray}
With the use of the relation (\ref{eq:entropy1}), 
the increase of the entropy (\ref{eq:increase}) is expressed as
\begin{eqnarray}
 \Delta={30 V_0 \over \pi^2 g_*T_C^3T_{\rm RH2}}\,, \label{eq:Delta-entropy}
\end{eqnarray}
where we have used $V_0=\rho(T_{\rm RH2})$.

The yield of the moduli $Y_\Phi$ after the flaton decay is given by
\begin{eqnarray}
 Y_{\Phi}={n_\Phi(T_{\rm RH2}) \over s(T_{\rm RH2})}={n_{\Phi}(T_C) \over s(T_C)\Delta}={n_\Phi(T_{\rm RH1}) \over s(T_{\rm RH1})\Delta}\,, \label{eq:yield}
\end{eqnarray}
where $n_\Phi$ is the number density of the moduli particles,
and we have assumed no entropy production before the end of the thermal inflation. 
Since the moduli particles are non-relativistic in this era, $n_\Phi$ 
at a certain temperature is represented by
\begin{eqnarray}
 n_\Phi={1 \over m_\Phi}\rho_\Phi\,,\label{eq:n-Phi}
\end{eqnarray}
where $\rho_\Phi$ is the energy density of the moduli.
The energy density of the moduli at $T_{\rm RH2}$
is produced by moduli oscillation after the primordial inflation:
\begin{eqnarray}
 \rho_\Phi={1 \over 2}m_\Phi^2 \Phi_0^2\,, \label{eq:moduli-o}
\end{eqnarray}
where  
$\Phi_0$ is the amplitude of the moduli fields.
During the moduli oscillation, the Universe is
in the matter-dominated era and therefore we have
\begin{eqnarray}
 \rho_{\Phi}(T_{\rm RH1}) = \rho_\Phi\left(a_{\rm osc} \over a(T_{\rm RH1})\right)^3
= \rho_\Phi\left(H(T_{\rm RH1}) \over H_{\rm osc}\right)^2\,, \label{eq:rho-phi}
\end{eqnarray}
where $a_{\rm osc}$ and $H_{\rm osc}$ are the scale factor and the Hubble parameter
when the moduli oscillation starts, and $a(T_{\rm RH1})$ and $H(T_{\rm RH1})$ are the ones at the reheating by the primordial
inflation.
Since the moduli oscillation starts when $H(T_0)\simeq m_{\Phi}$,
we express the moduli number density as 
\begin{eqnarray}
 n_\Phi(T_{\rm RH1})={1 \over 2m_\Phi}\Phi_0^2 H(T_{\rm RH1})^2\, , \label{eq:nphi}
\end{eqnarray}
from the expressions (\ref{eq:n-Phi}), (\ref{eq:moduli-o}) and (\ref{eq:rho-phi}).
The entropy density $s(T_{\rm RH1})$ in the denominator in (\ref{eq:yield}) is evaluated as
\begin{eqnarray}
 s(T_{\rm RH1})
={4 \over T_{\rm RH1}}\bar{M}_{\rm pl}^2H(T_{\rm RH1})^2\,,\label{eq:sRH1}
\end{eqnarray}
where we have used the relation
(\ref{eq:energy-density}) and the Friedmann equation
\begin{eqnarray}
 H^2(T_{\rm RH1})={\rho(T_{\rm RH1}) \over 3 \bar{M}_{\rm pl}^2}\,. \label{eq:freedman}
\end{eqnarray}
Substituting (\ref{eq:Delta-entropy}), (\ref{eq:nphi}) and
(\ref{eq:sRH1}) into (\ref{eq:yield}), 
we obtain the yield of the moduli:
\begin{eqnarray}
 Y_\Phi&=& {\pi^2 g_*\over 240}{T_{\rm RH1}T_{\rm RH2}T_C^3 \over m_\Phi V_0} \left(\Phi_0 \over \bar{M}_{\rm pl}\right)^2\,. \label{eq:yield-L}
\end{eqnarray}
With the use of the specific expressions of $T_{\rm RH2}$, $T_{\rm  C}$
and $V_0$ given in (\ref{eq:R-RH2}), (\ref{eq:tc}) and (\ref{eq:ph})
together with the decay width (\ref{eq:gamma1}), 
we have
\begin{eqnarray} 
Y_\Phi&=&{9\pi \over g^{3/2}}\left(g_* \over 10\right)^{3/4} {\lambda^{3/4}T_{\rm RH1} m_\phi \over m_\Phi \bar{M}_{\rm pl}} \nonumber\\
 &\simeq &{1.1 \times 10^{-6}\times  \lambda}^{3/4} \left(T_{\rm RH1} \over 10^{10}{\rm GeV}\right) \left(1 {\rm TeV} \over m_\Phi\right)\left( m_{\phi} \over 1 {\rm TeV}\right) \left(\Phi_0 \over \bar{M}_{\rm pl} \right)^2\,, \label{eq:yield-ex}
\end{eqnarray}
where we have chosen $g_*=200$ and $g=1$. It is natural that the moduli mass is the same order as the soft SUSY breaking  mass and the moduli amplitude is assumed to be the reduced Planck scale. Note that it is not necessary that $T_{\rm RH1}<10^6$ GeV to solve the gravitino problem since it can be solved after the thermal inflation as well. The moduli problem is solved if the yield satisfies the constraint \cite{Ellis:1990nb}
\begin{eqnarray}
 Y_\Phi<10^{-13}\,, \label{eq:condition}
\end{eqnarray}
which leads to an upper bound on $\lambda$ as
\begin{eqnarray}
 \lambda \lesssim 4.0 \times 10^{-10}\,, \label{eq:condition-lambda}
\end{eqnarray}
for $T_{\rm RH1}=10^{10}$ GeV, $m_{\Phi} = m_{\phi} = 1$ TeV, for example. 
Taking $\lambda=10^{-11}$ as a conservative value,
the reheating temperature $T_{\rm RH2}$ in (\ref{eq:R-RH2}) turns out to be
\begin{eqnarray}
 T_{\rm RH2}\simeq 164~{\rm MeV}\,. \label{eq:RH2-value}
\end{eqnarray}

\noindent{\bf Scenario 2}\\
Next we consider the scenario that the moduli oscillation starts after the
reheating by the primordial inflation (see Fig. \ref{fig:TH}). When the
moduli oscillation starts at $H\simeq m_\Phi$, the energy density of the
radiation becomes the same order of the energy density of moduli
(\ref{eq:moduli-o}) with $\Phi_0 \simeq \bar{M}_{\text{pl}}$.
With this observation, we find the temperature when the moduli oscillation starts
\begin{eqnarray}
 T_0=\left(15 \over \pi^2 g_*\right)^{1/4}\sqrt{m_\Phi \Phi_0}\,. \label{eq:osc-temp}
\end{eqnarray}
Considering that there is no entropy production until
the flaton decay, the yield of the moduli after the flaton decay is written as
\begin{eqnarray}
 Y_\Phi={n_\Phi(T_{\rm RH2}) \over s(T_{\rm RH2})}={n_\Phi(T_0) \over s(T_{0})\Delta}\,, \label{eq:yield2}
\end{eqnarray}
where the increase of entropy density $\Delta$ is the same as in
(\ref{eq:Delta-entropy}) since there is no entropy production during $T_0$ and $T_C$.

Substituting (\ref{eq:entropy1}), (\ref{eq:Delta-entropy}) and (\ref{eq:moduli-o}) into (\ref{eq:yield2}), we have
\begin{eqnarray}
 Y_\Phi&=&{3 \over 8}\left(\pi^2 g_* \over 15\right)^{3/4}{\Phi_0^{1/2}
  T_C^3T_{\rm RH2} \over m_\Phi^{1/2}V_0}\,.\label{eq:yield-s2}
\end{eqnarray}
Combining this result with (\ref{eq:tc}) and (\ref{eq:R-RH2}) we obtain
\begin{eqnarray}
Y_\Phi &=& 27\times {6^{1/4}g_*^{1/2} \over g^{3/2}}\sqrt{\pi \over 5}{\lambda^{3/4}m_\phi \Phi_0^{1/2}\over m_\Phi^{1/2}\bar{M}_{\rm pl}} \nonumber\\ 
&\simeq &6.2\times 10^{-6}\lambda^{3/4}\left({m_\phi \over 1 {\rm TeV}}\right)\left({1 {\rm TeV} \over m_\Phi}\right)^{1/2}\left({\Phi_0 \over \bar{M}_{\rm pl}}\right)^{1/2}\,,
\end{eqnarray}
where we have chosen $g_*=200$ and $g=1$. The condition
(\ref{eq:condition}) leads to
\begin{eqnarray}
 \lambda \lesssim 4.0\times 10^{-11}\,.
\end{eqnarray}
The reheating temperature after the flaton decay is given as
\begin{eqnarray}
 T_{\rm RH2}\simeq 92~{\rm MeV}\,,
\label{eq:RH2-value2}
\end{eqnarray}
for a conservative value $\lambda=10^{-12}$.

In both scenarios, the reheating temperature is sufficiently high to realize the Big-Bang nucleosynthesis. However, since the thermal inflation dilutes the primordial baryon asymmetry, we need to consider baryogenesis after the thermal inflation. A simple baryogenesis such as the thermal leptogenesis \cite{FY} and the electroweak baryogenesis (for review, see, e.g. \cite{Cohen:1993nk}) can not be operative with 
such the low reheating temperatures in (\ref{eq:RH2-value}) and \eqref{eq:RH2-value2}.
In order for the thermal leptogenesis to work, 
$T_{\rm RH2}\gtrsim 10^3$ GeV is necessary \cite{Dev:2014laa}. On the other hand, the Affleck-Dine mechanism \cite{AD, DRT} could be implemented with (\ref{eq:RH2-value}), as 
has been studied in models of the thermal inflation \cite{SKY,HMY}. 

We mentioned that the negative mass squared for the flaton field in (\ref{eq:potential}) can be realized by 
the renormalization group effect \cite{Murayama:1992dj}. For instance, assume that the flaton mass squared is positive at a scale where the primordial inflation ends and the flaton couples to a scalar field through the Yukawa interaction in the superpotential.
In a certain condition, the Yukawa interaction drives the flaton mass squared negative.
However, in order to realize this, it is likely that the Yukawa coupling beyond the perturbative regime is necessary \cite{Murayama:1992dj}. 
In the next section, we will propose a simple
scenario to  terminate the thermal inflation. We will also show that in
a proposed scenario the reheating temperature $T_{\rm RH2}$ can be much
larger than $10^3$ GeV, which makes it possible to implement the thermal
leptogenesis.

%
%
\section{Thermal inflation with chemical potential}
In this section, we introduce the chemical potential for the flaton in the thermal
inflation scenario and study its effect.
The existence of the chemical potential means that the flaton is 
dense at a vacuum realized after the end of the thermal inflation. 
It has been shown that there exists such a vacuum with non-zero chemical potential in ${\cal N}=1$ supersymmetric QCD \cite{HaLaMu}.

Considering the moduli problem stems from the superstring theories, 
it is natural to embed a model
in the supersymmetry framework.
In the following, we consider a supersymmetric model where the flaton
field $\phi$ and a scalar field $X$, both of which are complex,
are realized as the lowest components of $\mathcal{N} = 1$ chiral
superfields. 
We begin with the following tree level scalar potential of these fields:

\begin{eqnarray}
 V=V_0+{\lambda \over \bar{M}_{\rm pl}^2}|\phi|^6+m_{X
  0}^2|X|^2+g|\phi|^2|X|^2\,.
\label{eq:complex_potential}
\end{eqnarray}
The potential \eqref{eq:complex_potential} exhibits the $U(1)_c$ global
symmetry under the transformation $\phi\rightarrow e^{i\alpha}\phi$. 
Here the constant $\alpha$ is the $U(1)_c$ charge.
The chemical potential is
introduced by gauging the $U(1)_c$ global
symmetry for the flaton
\cite{Actor,HaLaMu}. The spacetime derivative is replaced with the gauge covariant
derivative $D_\mu=\partial_\mu+i \alpha A_\mu$, where
$A_\mu$ is a non-dynamical gauge field. The gauge field has the vacuum
expectation value only in the zeroth component $\langle A_\mu \rangle=(i\mu, {\bf 0})$. 
Note that the field $X$, which is in the thermal equilibrium, 
is neutral under the $U(1)_c$ transformation.
We also note that although the complex scalar field $\phi$ leads to a multi-flaton model, we can
always rotate away the imaginary (real) part of $\phi$ during the
inflation by the $U(1)_c$ transformation. Therefore the inflation
dynamics does not change from single flaton models.

The partition function with the non-zero temperature and the chemical potential is written as
\begin{eqnarray}
Z &=& {\rm Tr}e^{-\beta(H-\mu{\cal N})} \nonumber \\
 &=&\int_{\Phi_i(\tau)=\Phi_i(\tau+\beta)} \prod_i
{\cal D}\Phi_i{\cal D}\Phi_i^\dagger\; e^{-\int_0^\beta d\tau \int d^3x (D_0 \phi D_0 \phi^\dagger+\partial_0 X \partial_0 X^\dagger+\sum_i\vec{\nabla} \Phi_i \vec{\nabla}\Phi_i^\dagger+V)}\,,
\end{eqnarray}
where ${\cal N}$ is the Noether charge of the $U(1)_c$ symmetry, $\Phi_i=(\phi,X)$ and $D_0={\partial \over \partial \tau}-\mu$ with the unit $U(1)_c$ charge $\alpha=1$. The thermal effective potential for the flaton $\phi$
after the primordial inflation is obtained by calculating the thermal 1-loop correction of $X$:
\begin{eqnarray}
V_{\rm eff}=V_0-\mu^2|\phi|^2+{\lambda \over \bar{M}_{\rm pl}^2}|\phi|^6+\int {d^3k \over (2\pi)^3}{\omega_k \over 2}
+{1 \over \beta}\int {d^3k \over (2\pi)^3} \log(1-e^{-\beta \omega_k})\,. \label{pc}
\end{eqnarray}
Note that the chemical potential yields a negative mass squared for the flaton.
This potential has the same form with (\ref{eff-pot-T}) when $\mu$ is
replaced with $m_{\phi_0}$. However, it should be emphasized that $\mu$
can be in general any value, while $m_{\phi 0}\simeq {\cal O}(1)$ TeV in the standard thermal inflation scenario
since $m_{\phi 0}$ is considered to be caused by SUSY breaking. 
The fourth term in (\ref{pc}) is the Coleman-Weinberg
potential, which we will omit in the following discussion. 
The fifth term is the thermal effective potential with non-zero chemical potential, 
where $\omega_k$ is given in (\ref{eq:omega}) with (\ref{eq:mass-matrix}).

We study the thermal inflation with this potential and how the moduli
problem is solved in an analytic way. Performing 
the high-temperature expansion, we have
\begin{eqnarray}
V_{\rm eff}=V_0 - {\pi^2 T^4 \over 45} +{T^2 \over 12}m_{X
 0}^2+\left(-\mu^2+{g T^2 \over 12}\right)|\phi_c|^2 + {\lambda \over
 \bar{M}_{\rm pl}^2} |\phi_c|^6+\cdots\,. \label{eq:eff-potential2}
\end{eqnarray}
When the coefficient of $|\phi_c|^2$ is positive, the potential minimum
is at the origin for $\phi_c$ and the thermal inflation takes place.   
According to the expansion of the Universe, the temperature is decreasing, and 
the thermal inflation eventually ends at the critical temperature given by
\begin{eqnarray}
 T_C=2\mu \sqrt{3 \over g}\,. \label{eq:tc2}
\end{eqnarray}
Below this temperature, the flaton rolls down to the vacuum 
which is determined by the extreme condition
\begin{eqnarray}
 {\partial^2 V \over \partial \phi \partial \phi^\dagger}{\Big |}_{X=0, T=0}=0\,.
\end{eqnarray}
From this condition, we have
\begin{eqnarray}
\phi_c=(3\lambda)^{-1/4}\sqrt{\mu \bar{M}_{\rm pl}} \equiv M_c \,. \label{eq:vac2}
\end{eqnarray}
The flaton mass at the vacuum is given as
\begin{eqnarray}
 m_\phi^2={\partial^2 V \over \partial \phi \partial \phi^\dagger}{\Big |}_{T=0, \phi_c=M_c}=\mu^2\,. \label{eq:fmass}
\end{eqnarray}
The potential energy $V_0$ is determined so that the scalar potential is vanishing at the vacuum:
\begin{eqnarray}
 V_0 ={2 \over 3\sqrt{3\lambda}}\mu^3\bar{M}_{\rm pl}={2 \over 3}\mu^2 M_c^2\,. \label{eq:zero-energy}
\end{eqnarray}

The flaton oscillates around the vacuum 
and the thermalization then occurs. In order to evaluate the reheating temperature, we need to specify the interaction of the flaton with 
the Standard Model fields. The interaction considered in (\ref{eq:int}) cannot be employed
since this does not preserve the $U(1)_c$ symmetry related
to the chemical potential. 
Instead, we consider the following $U(1)_c$ preserving interaction 
(see Appendix B for the derivation):
\begin{eqnarray}
 {\cal L}_{\rm int}= \sum_{a=1}^3c_a {M_c \over \bar{M}_{\rm pl}^2}\chi\left(-{1 \over 4}F^{a\mu\nu}F^a_{\mu\nu}\right)\,, \label{int-final}
\end{eqnarray}
where $\chi\equiv {\rm Re}(\phi)$, $c_a(a=1,2,3)$ is a constant and
$F_{\mu\nu}^a$ is the gauge field strength. Here the index $a=1,2,3$
corresponds to the Standard Model gauge groups, $SU(3)\times SU(2)_L\times U(1)_Y$. 
The partial decay widths of $\chi$ into the Standard Model gauge bosons are calculated to be \cite{Itoh:2006fv}
\begin{eqnarray}
 \Gamma(\chi\rightarrow gg)&=&{c_3^2 \over 2\pi}\left(M_c \over \bar{M}_{\rm pl}^2\right)^2\mu^3\,, \\
 \Gamma(\chi\rightarrow \gamma\gamma)&=&{(c_1\cos^2\theta_W+c_2\sin^2\theta_W)^2 \over 16\pi}\left(M_c\over \bar{M}_{\rm pl}^2 \right)^2\mu^3\,,\\
 \Gamma(\chi\rightarrow ZZ)&=&{(c_1\cos^2\theta_W+c_2\sin^2\theta_W)^2 \over 128\pi}\left(M_c\over \bar{M}_{\rm pl}^2 \right)^2\mu^3\beta_Z(3+2\beta_Z^2+3\beta_Z^4)\,,\\
  \Gamma(\chi\rightarrow WW)&=&{c_2^2 \over 64\pi}\left(M_c\over \bar{M}_{\rm pl}^2 \right)^2\mu^3\beta_W(3+2\beta_W^2+3\beta_W^4)\,,\\
 \Gamma(\chi\rightarrow \gamma Z)&=&{(c_1-c_2)^2\sin^2\theta_W\cos^2\theta_W \over 8\pi}\left(M_c\over \bar{M}_{\rm pl}^2 \right)^2\mu^3\left(1-{m_Z^2 \over \mu^2}\right)^3\,,
\end{eqnarray}
where $\theta_W$ is the weak mixing
angle and $\beta_Z=\sqrt{1-4m_Z^2/\mu^2}$ and
$\beta_W=\sqrt{1-4m_W^2/\mu^2}$.
Here $m_Z$, $m_W$ are masses of the Z and W bosons.
With the use of these decay widths,
the reheating temperature is obtained to be
\begin{eqnarray}
 T_{\rm RH2} \simeq  \left(90 \over \pi^2 g_*\right)^{1\over 4}\sqrt{3 \over 4\pi}{\mu^2 \over (3\lambda)^{1/4}\bar{M}_{\rm pl}}\,,\label{eq:RH2-2}
\end{eqnarray}
where we have chosen $c_1=c_2=c_3=1$, for simplicity, and have put $\beta_Z\simeq 1$ and $\beta_W\simeq 1$ since we assume $m_Z, ~m_W \ll\mu$.

We now evaluate the yield (\ref{eq:yield}) through the flaton decay in
two scenarios: Moduli starts to oscillate before (Scenario 1) and after
(Scenario 2) the reheating by the primordial inflation. The chemical
potential just plays a role of the flaton mass and  does not affect
the derivation of the yield from (\ref{eq:increase}) to
(\ref{eq:yield-L}) for Scenario 1 and from \eqref{eq:osc-temp}
to (\ref{eq:yield-s2}) for Scenario 2 in the previous section. 
Therefore we have the same formula for the yield as (\ref{eq:yield-L})
for Scenario 1 and (\ref{eq:yield-s2}) for Scenario 2. 

For Scenario 1, substituting (\ref{eq:tc2}), (\ref{eq:zero-energy}) and (\ref{eq:RH2-2}) into (\ref{eq:yield-L}), we find
\begin{eqnarray}
Y_\Phi &=& {9 \pi\over 4g^{3/2}}\left(3g_* \over 10\right)^{3/4}{\lambda^{1/4} \mu^2 \over \bar{M}_{\rm pl}^2}\times 10^7 \left(T_{\rm RH1} \over 10^{10} {\rm GeV}\right)\left(1 {\rm TeV} \over m_\Phi\right) \left(\Phi_0 \over \bar{M}_{\rm pl}\right)^2\,\nonumber \\
&\simeq & 1.5 \times 10^{9}\times {\lambda^{1/4}\mu^2 \over \bar{M}_{\rm pl}^2}\left(T_{\rm RH1} \over 10^{10} {\rm GeV}\right)\left(1 {\rm TeV} \over m_\Phi\right)\left(\Phi_0 \over \bar{M}_{\rm pl} \right)^2\,, \label{eq:yield-ex2}
\end{eqnarray}
where we have taken $g=1$ and $g_*=200$.
The moduli problem is resolved when the yield \eqref{eq:yield-ex2}
satisfies the condition \eqref{eq:condition}.
In other words, $\lambda$ and $\mu$ should satisfy the following condition
\begin{eqnarray}
 1.5 \times {\lambda^{1/4}\mu^2 \over \bar{M}_{\rm pl}^2}\lesssim 10^{-22}\,. \label{eq:sol}
\end{eqnarray}

For Scenario 2, repeating the same derivation for (\ref{eq:yield-s2}) we obtain
\begin{eqnarray}
Y_\Phi &=& {81 \over 2\times 2^{3/4}(10g)^{3/2}}\left(\pi g_* \over 5\right)^{1/2}{\lambda^{1/4}\mu^2 \over \bar{M}_{\rm pl}^{3/2}}\left(1 {\rm TeV} \over m_\Phi\right)^{1 \over 2}\left(\Phi_0 \over \bar{M}_{\rm pl} \right)^{1\over 2}\, \nonumber \\
&\simeq &2.7\times 10^2 \times {\lambda^{1/4}\mu^2 \over \bar{M}_{\rm pl}^{3/2}}\left(1 {\rm TeV} \over m_\Phi\right)^{1 \over 2}\left(\Phi_0 \over \bar{M}_{\rm pl} \right)^{1\over 2}\,.
\label{eq:yield-ex3}
\end{eqnarray}
This expression with the condition (\ref{eq:condition}) leads to
\begin{eqnarray}
2.7 \times {\lambda^{1/4}\mu^2 \over \bar{M}_{\rm pl}^{3/2}}\lesssim10^{-15}\,. \label{eq:sol2}
\end{eqnarray}

Allowed values for $\lambda$ and $\mu$ for the conditions (\ref{eq:sol}) and (\ref{eq:sol2}) determine the reheating temperature (\ref{eq:RH2-2}). 
We may require the reheating temperature high enough
to realize the thermal leptogenesis \cite{FY}, such as $T_{\rm RH2}\gtrsim 1$ TeV. 
On the other hand, the consistency of our discussion requires $T_C>T_{\rm RH2}$, which leads to
\begin{eqnarray}
 \lambda>{1 \over 3}\left(90 \over \pi^2 g_*\right)\left(g \over 32\pi\right)^2 {\mu^2 \over \bar{M}_{\rm pl}^2}\,, \label{eq:TRH2TC}
\end{eqnarray}
where we have used (\ref{eq:tc2}) and (\ref{eq:RH2-2}). 
The coupling constant $\lambda$ and the chemical potential 
$\mu$ are also constrained from the condition that the vacuum
expectation value of the flaton should be less than the Planck scale 
$M_c<\bar{M}_{\rm pl}$. This results in the following condition:
\begin{align}
\lambda > {\mu^2  \over 3 \bar{M}_{\rm pl}^2}\,. \label{eq:pl}
\end{align}

\begin{figure}[tb]
\centering
\includegraphics[scale=0.19]{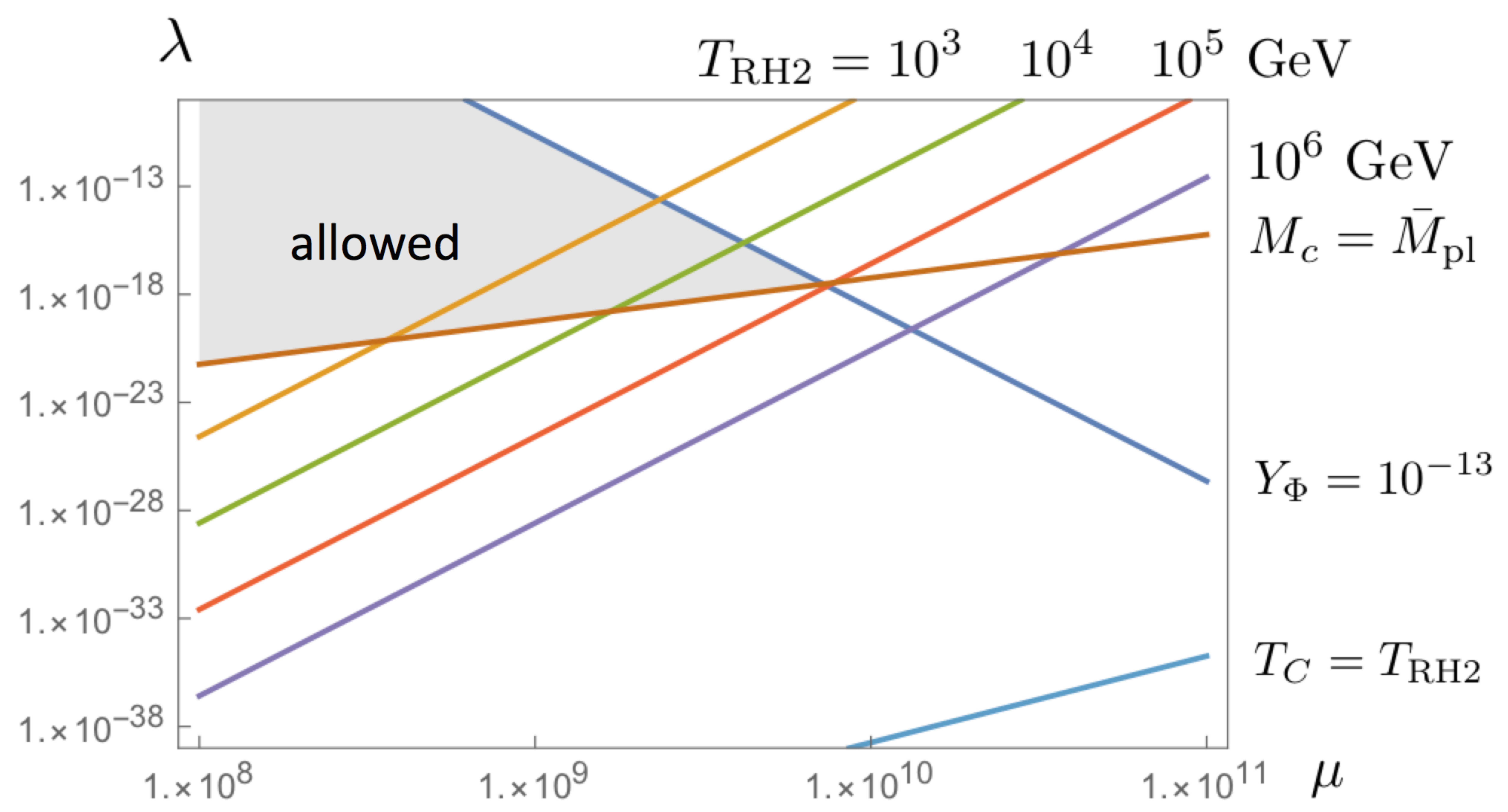}
 \caption{Allowed parameter region for $\lambda$ and $\mu$ in Scenario 1.
Here we have taken $g_*=200$ and $g=1$.}
\label{scenario1}
\end{figure}

Fig. \ref{scenario1} shows the parameter region for Scenario 1 that
satisfies (\ref{eq:sol}), (\ref{eq:TRH2TC}) and (\ref{eq:pl}) together
with the lines corresponding to
the reheating temperature $T_{\rm RH2}=10^3, 10^4,
10^5$ and $10^6$ GeV. Here we have taken $g_*=200$ and $g=1$. 
We can see that the condition (\ref{eq:pl}) is stronger than (\ref{eq:TRH2TC}). Indeed,
(\ref{eq:pl}) with (\ref{eq:sol}) sets the upper bound 
on the chemical potential as $\mu \lesssim 7.2\times 10^9$ GeV. Considering that the
thermal leptogenesis is operative at least for $T_{\rm RH2}\gtrsim 10^3$
GeV along with (\ref{eq:pl}), we find the lower bound 
on $\lambda$ as $\lambda\gtrsim 7.5\times 10^{-21}$. It is possible to increase $T_{\rm
RH2}$ up to around $9.0\times 10^4$ GeV, beyond which the vacuum
expectation value $M_c$ is larger than the Planck scale.

%
%

\begin{figure}[tb]
\centering
\includegraphics[scale=0.2]{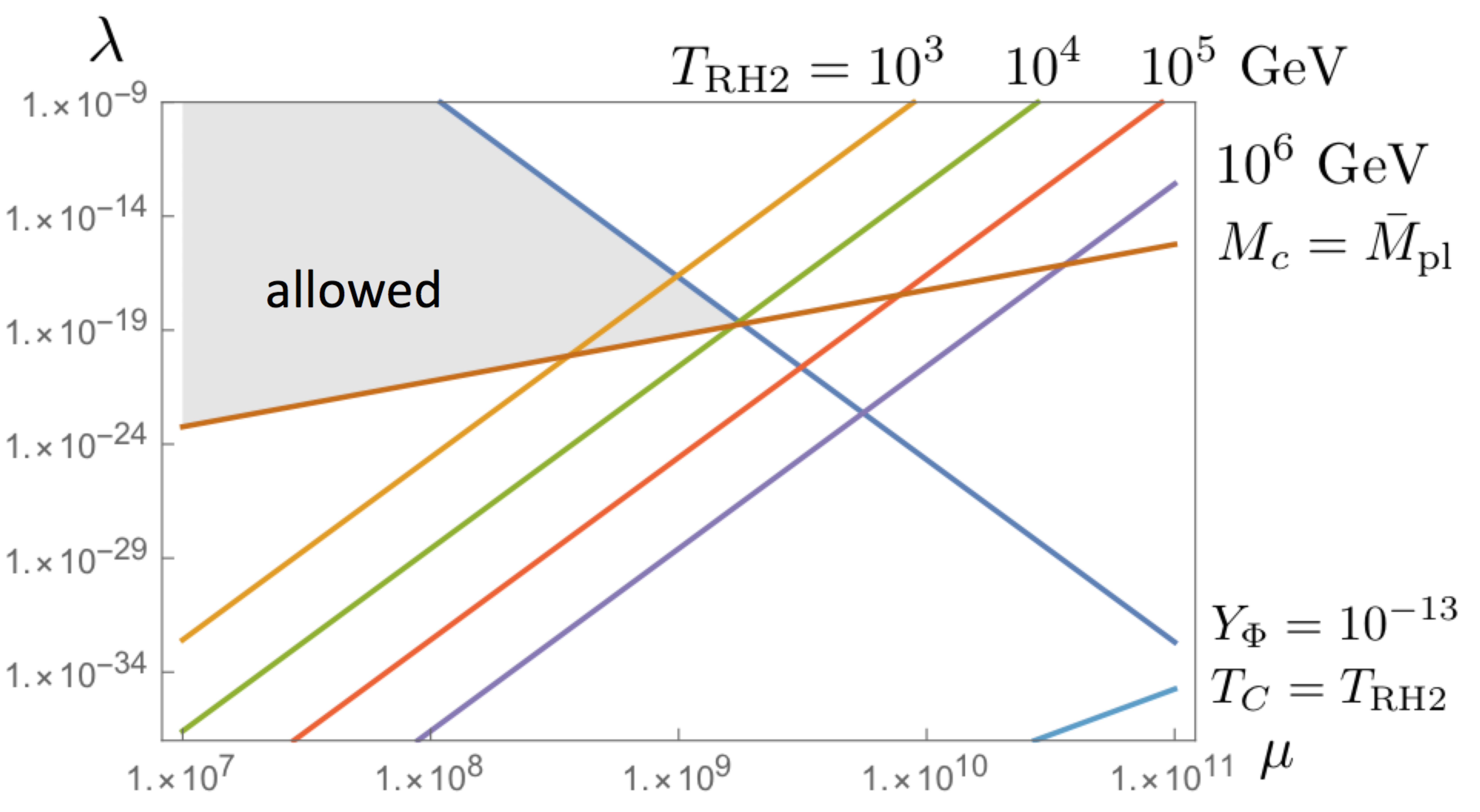}
 \caption{
Same as Fig. \ref{scenario1} but for Scenario 2.
}
\label{scenario2}
\end{figure}

A similar figure for Scenario 2 is shown in Fig. \ref{scenario2}. The
upper bound on the chemical potential is given as $\mu \lesssim
1.5\times 10^9$ GeV and the lower bound 
on $\lambda$ such that the thermal leptogenesis is operative is found to
be $\lambda \gtrsim 7.5\times 10^{-21}$. The reheating temperature can be taken up to $8.6\times 10^3$
GeV, which is smaller than the one in Scenario 1.

%
%

It should be emphasized that in the standard thermal inflation scenario,
$T_{\rm RH2}$ cannot be large enough to implement the baryogenesis
scenario except for the Affleck-Dine mechanism. The reheating temperature
(\ref{eq:R-RH2}) is proportional to $\lambda^{1\over 4}$ and the flaton
mass. Recalling that 
$m_{\phi}\simeq 1$ TeV
and $\lambda$ should be
small enough to satisfy (\ref{eq:condition-lambda}), we see that $T_{\rm
RH2}$ in the standard thermal inflation scenario
is at most $\mathcal{O} (100)$ MeV. 
However, in our scenario the reheating temperature (\ref{eq:RH2-2}) is proportional to $\lambda^{-1/4}$ and $\mu^2$. Since $\mu$ is taken to be larger than $1$ TeV and in addition $\lambda$ can be taken to be small to satisfy (\ref{eq:sol}) and (\ref{eq:sol2}) (but it is 
constrained by (\ref{eq:pl})), one can realize a reheating temperature high enough to implement the thermal leptogenesis.

%
%
\section{Conclusion}
In this paper, we have studied the models of the thermal inflation with
the flaton chemical potential which is implemented naturally 
by the VEV of the zeroth component of the $U(1)_c$ (non-dynamical) gauge
field. This leads to the negative mass squared of the flaton. 
On the other hand, in the standard thermal inflation, a negative mass
squared of ${\cal O}(1)$ TeV,
which is the soft SUSY breaking scale, can be realized by the renormalization group flow
with a large coupling constant (most likely to be in the non-perturbative regime); 
otherwise, it is introduced by hand.
We have evaluated the yield of the moduli in two scenarios: Moduli
field start to oscillate before (Scenario 1) and after (Scenario 2) the
reheating by the primordial inflation. In both scenarios, the yield
depends on $\lambda$ being the coefficient of the sixth order term of
the flaton potential and the chemical potential $\mu$. We have found the
allowed parameter region in the $(\lambda, \mu)$-plane, in which after
the thermal inflation
the reheating temperature can be high enough for the thermal
leptogenesis to be operative. This is in sharp contrast to 
the standard thermal inflation, in which the reheating temperature is at
most ${\cal O}(100)$ MeV. 

In this work we have introduced the flaton chemical potential as a free
parameter. It is worth investigating a possible origin of the chemical
potential in the framework of superstring theories. It is also interesting to
consider a possibility to relate the global $U(1)_c$ to the baryon or
the lepton numbers in the Standard Model.

%
%
\appendix
\section*{Appendices}
\renewcommand{\thesubsection}{\Alph{subsection}}
\subsection{Thermal 1-loop correction}
 \renewcommand{\theequation}{A-\arabic{equation}}
 \setcounter{equation}{0}  
\label{appendixA}
In this appendix,
we give a sketch of the derivations for (\ref{eff-pot-T}) and (\ref{pc}).
For the details, consult the references \cite{Bernard:1974bq, Dolan:1973qd}.

At the 1-loop level, the correction of the effective potential
 from the thermal effect for a real scalar field
is given by the determinant,
\begin{eqnarray}
\log \left(\det \left(\partial^2 + m^2 \right) \right)^{1/2} &=&
\frac{1}{2} {\rm tr} \; \log \left(\partial^2 + m^2 \right)
 \nonumber \\
&=& \frac{1}{2\beta} \sum_{n=- \infty}^{+\infty} \int \frac{d^3 k}{(2 \pi)^3} \ 
\log \left( (2 \pi \beta^{-1} n)^2 + \omega_k \right),
\label{TrLog}
\end{eqnarray}
in the absence of the chemical potential.
After formally differentiating (\ref{TrLog}) with respect to $\omega_k$,
we can sum over the discrete momentum $ 2 \pi \beta^{-1} n$,
\begin{eqnarray}
\sum_{n=- \infty}^{+\infty} \frac{ \partial}{\partial \omega_k}
\log \left( (2\pi \beta^{-1} n)^2 + \omega_k^2 \right) &=&
\sum_{n=- \infty}^{+\infty}
\frac{2 \omega_k}{(2\pi \beta^{-1} n)^2 + \omega_k^2 }
 \nonumber \\
&=& \sum_{n=- \infty}^{+\infty}  \left(
 \frac{1}{\omega_k - i(2\pi \beta^{-1} n)}
+  \frac{1}{\omega_k + i(2\pi \beta^{-1} n)} \right) \nonumber \\
&=& \beta \coth \left( \frac{\beta \omega_k}{2} \right).
\label{SumCot}
\end{eqnarray}
Here we use the partial fraction expansion formula,
\begin{equation}
\pi \coth(\pi x ) = \sum_{n=- \infty}^{+\infty} \frac{1}{x+in}.
\end{equation}
Integrating (\ref{SumCot}) with $\omega_k$,
we obtain 
\begin{eqnarray}
\log \left(\det  \left(\partial^2 + m^2 \right)\right)^{1/2} &\simeq&
 \frac{1}{\beta} \int \frac{ d^3 k}{(2 \pi)^3} \  \log \left|
 \sinh \left( \frac{\beta \omega_k}{2} \right) \right|
 \nonumber \\
&\simeq&  \int \frac{d^3 k}{(2\pi)^3} \left( \frac{\omega_k}{2}
 + \frac{1}{\beta} \log \left| 1 - e^{-\beta \omega_k}
\right| \right),
\label{DetLogResult}
\end{eqnarray}
up to an irrelevant constant.
In the case of a complex scalar, the correction is twice of that of a real scalar.

\subsection{Interaction terms of the flaton with the Standard Model sector}
 \renewcommand{\theequation}{B-\arabic{equation}}
 \setcounter{equation}{0}  
\label{appendixB}
We consider the following 
higher dimensional term invariant under the $U(1)_c$ transformation for the flaton, $\phi\rightarrow e^{i\alpha}\phi$ associated with the chemical potential.
\begin{eqnarray}
{\cal L}_{\rm int}={1 \over 4}\int d^4 \theta \sum_{a=1}^3 c_a {\Phi^\dagger \Phi \over \bar{M}_{\rm pl}^2} \left(W^{a\alpha} W^a_\alpha \delta^2(\bar{\theta})+h.c.\right)\,, \label{int-eq}
\end{eqnarray}
where $\Phi$ is a chiral superfield associated with the flaton
and $W_\alpha^a$ is a superfield strength with the index $a=1,2,3$ corresponding to
the Standard Model gauge groups $SU(3) \times SU(2)_L \times U(1)_Y$. 
In order to consider the interaction at the vacuum $\langle \Phi \rangle = M_c$, we substitute a shift  
\begin{eqnarray}
 \Phi \rightarrow M_c + \Phi\,,
\end{eqnarray}
into (\ref{int-eq}) and pick up the following three-point vertex part:
\begin{eqnarray}
 {\cal L}_{\rm int}\supset {M_c \over 4\bar{M}_{\rm pl}^2}\int d^4\theta  \sum_{a=1}^3 c_a(\Phi+\Phi^\dagger) (W^{\alpha a}W^a_\alpha \delta^2(\bar{\theta})+h.c.)\,. \label{int2}
\end{eqnarray}
Since we are interested in the flaton decay, we focus on the scalar part of the flaton superfield, $\phi=\Phi|_{\theta=0}$ in (\ref{int2}):
\begin{eqnarray}
 {\cal L}_{\rm int}\supset {M_c \over \bar{M}_{\rm
  pl}^2}\chi\sum_{a=1}^3 c_a \left(-{1 \over 4}F_{\mu\nu}^aF^{\mu\nu
							  a}-i\lambda^a\sigma^\mu \partial_\mu
							  \bar{\lambda}^a\right)\,, 
\end{eqnarray}
where $\chi\equiv {\rm Re}(\phi)$, $F_{\mu\nu}^a$ and $\lambda^a$ are
the field strength and the gaugino, respectively. This interaction leads
to the decays $\chi\rightarrow A_\mu A_\nu$ and $\chi\rightarrow
\lambda\bar{\lambda}$. The decay widths are obtained as
$\Gamma(\chi\rightarrow A_\mu A_\nu)\propto (M_c/\bar{M}_{\rm
pl}^2)^2\mu^3$ and $\Gamma(\chi\rightarrow \lambda\bar{\lambda})\propto
(M_c/\bar{M}_{\rm pl}^2)^2m_{\lambda}^2\mu$, where $m_\lambda \simeq 1 $ TeV 
is the gaugino mass.
Since we take $\mu \gg 1$ TeV in our scenario
(see Figs. \ref{scenario1} and \ref{scenario2}), the flaton
mainly decays to the Standard Model gauge bosons.

%
%
\subsection*{Acknowledgments}
This work is supported in part by the Japan Society for the Promotion for
Science Grant-in-Aid for Scientific Research (KAKENHI) Grant Numbers 
25400280 (M.A.), the United States Department of Energy (DE-SC001368)(N.O.) 
and Kitasato University Research Grant for Young Researchers (S.S.).

%
%

\end{document}